\documentclass[twocolumn,prb,aps,floatfix,showpacs]{revtex4-1}

\usepackage{graphics}%
\usepackage{dcolumn}

\begin{document}
\bibliographystyle{apsrev}
\title[interstitials in GaAs]{First-principles study of As 
interstitials in GaAs: Convergence, relaxation, and formation energy}

\author{J. T. Schick}
\email[]{joseph.schick@villanova.edu}
\affiliation{Physics Department, Villanova University, Villanova, 
PA 19085, USA}
\author{C. G. Morgan}
\author{P. Papoulias} 
\affiliation{Department of Physics and Astronomy, Wayne State University, 
Detroit, MI 48202, USA}

\date{August 19, 2002}

\begin{abstract}

Convergence of density-functional supercell 
calculations for defect formation energies, charge 
transition levels, localized defect state properties,
and defect atomic structure 
and relaxation is investigated using the arsenic split 
interstitial in GaAs as an example.  
Supercells containing up to 217 atoms and a variety 
of {\bf k}-space sampling schemes are considered.  
It is shown that a good description of the 
localized defect state dispersion and charge state 
transition levels requires at least  
a 217-atom supercell, although the defect structure and atomic relaxations can 
be well converged in a 65-atom cell.
Formation energies are calculated for the As split interstitial, 
Ga vacancy, and 
As antisite defects in GaAs, taking into account the 
dependence upon chemical potential and Fermi energy.
It is found that equilibrium concentrations of As interstitials will 
be much lower than equilibrium concentrations of As antisites in 
As-rich, $n$-type or semi-insulating GaAs.
\end{abstract}


\maketitle

\section{Introduction}
Interstitials are the most complicated of the simple point defects, 
and the most elusive.  For example, even though arsenic interstitials must be 
created by irradiation of GaAs with sufficiently energetic particles, and they 
can subsequently be observed to recombine with arsenic vacancies when the 
sample is heated above $220^\circ$~C, 
isolated arsenic interstitials have not been 
observed directly in EPR, electrical, or 
optical experiments.\cite{Bourgoin1988}  

It has been argued based on a thorough analysis 
\cite{Hurle1999,Hurle1995a,Hurle1992} of a 
variety of experimental data including 
titration experiments \cite{Oda1992} and measurements of density and lattice 
parameter\cite{Bublik1973} that melt-grown GaAs is always As-rich unless 
the concentration of Ga in the melt is substantially greater than 50\%, 
and that this deviation from stoichiometry is due primarily to the creation 
of large numbers of As interstitials ($\mathrm{As}_i$) during growth. 
In particular, Hurle has
argued\cite{Hurle1999} that the measured deviation 
of the mass per unit cell as a
function of arsenic concentration in the melt must be explained by 
arsenic interstitials and/or arsenic vacancies, 
since the number of arsenic 
antisites which would be required to fit the data is unrealistically large 
(up to several percent), due to the small difference between the atomic 
masses of arsenic and gallium.  Hurle's work also contains an extensive 
thermodynamic analysis, including estimates 
of the mass action constants for the 
formation of all the neutral native point defects.  These estimates
are derived by fitting to a large quantity of experimental data 
on both doped and undoped GaAs, under the assumption that 
native defect and dopant concentrations are near equilibrium close to the 
melting point and during high temperature growth from 
the melt or from solution.  

In the high temperature growth regime, observations of defects  
tentatively described as high concentrations or diffuse `clouds' of 
arsenic interstitials have been reported 
in GaAs grown by the horizontal Bridgman and liquid-encapsulated Czochralski 
methods, based on X-ray diffuse 
scattering\cite{Charniy1994a,Charniy1992a,Charniy1992b} and 
quasi-forbidden X-ray reflection intensity measurements.\cite{Fujimoto1992}  
However, the atomic composition and microscopic 
structure of these defects cannot be unambiguously determined from these 
experiments.  

Gallium arsenide grown by arsenic-rich 
molecular beam epitaxy at low temperature
(LT GaAs) is a 
semi-insulating material with a host of potentially useful 
applications.\cite{Nolte1999}  
This material contains up to 1.5\% excess As,\cite{Yu1992}
which is accommodated by high concentrations of point defects 
in UN-annealed samples, and arsenic precipitates plus somewhat lower 
concentrations of point defects in annealed samples.
Concentrations of As antisites ($\mathrm{As_{Ga}}$) up to 
$10^{20}$~cm$^{-3}$ 
are observed in LT GaAs, as measured by electron 
paramagnetic resonance (EPR),\cite{Kaminska1989} near-infrared 
absorption (NIRA) and magnetic circular dichroism of 
absorption (MCDA),\cite{Liu1995} and scanning tunneling microscopy 
(STM).\cite{Feenstra1993}  Concentrations of Ga vacancies 
($V_{\mathrm{Ga}}$) up to $10^{18}$~cm$^{-3}$ are measured in 
LT GaAs by slow positron annihilation.\cite{Luysberg1998}  Ion 
channeling experiments have been interpreted as providing evidence 
for large concentrations of As interstitials in 
LT GaAs.\cite{Yu1992}
However, it was later pointed out that the observed high concentration 
of atoms in
the channel near the normal arsenic lattice sites could also be due to outward 
relaxation of the nearest neighbors of the 
As antisites.\cite{Liu1995,Staab2001} 

Within certain well defined limits 
of the growth parameters for LT GaAs, a linear correlation between 
the neutral $\mathrm{As_{Ga}}$
concentration and the lattice dilation has been 
found.\cite{Liu1995,Luysberg1998}  
It was therefore proposed 
that $\mathrm{As_{Ga}}$ are the dominant defects which determine the lattice 
expansion for growth within this regime. 
Staab {\em et al.}\ \cite{Staab2001} used a self-consistent 
density-functional-based tight-binding method 
to study the lattice distortion induced by 
point defects in As-rich GaAs, and concluded that only
$\mathrm{As_{Ga}}$ are necessary to understand the observed lattice 
expansion in the regime where the linear correlation is observed, and 
that if concentrations of isolated As$_i$ 
comparable to the measured concentrations of 
$\mathrm{As_{Ga}}$ were also present, the lattice 
expansion would be three times
greater than is experimentally observed.
However, Luysberg {\em et al.}\ have reported that when the As/Ga flux 
ratio is increased beyond a beam equivalent pressure (BEP) ratio of 20, 
there is a departure from the 
linear correlation between lattice dilation and antisite 
concentration.\cite{Luysberg1998}  It was pointed 
out that other defects must be present to account for the deviation from 
stoichiometry and the lattice expansion at high As/Ga 
flux ratios.\cite{Luysberg1998}

Nonequilibrium processes such as diffusion and compositional intermixing at 
interfaces can also be strongly affected by point 
defects that are present in high 
concentrations.  
Since the point defects which have been unambiguously documented as
present in high concentrations in LT GaAs, 
$\mathrm{As_{Ga}}$ and $V_\mathrm{{Ga}}$, 
occupy sites on the gallium sublattice, they cannot contribute directly to 
interdiffusion on the arsenic sublattice.  However, substantial concentrations 
of arsenic interstitials 
may affect interdiffusion on the arsenic sublattice.  
For example, an experimental study showing a positive 
dependence of GaAsP/GaAs and 
GaAsSb/GaAs interdiffusion on arsenic pressure has indicated that a 
kickout mechanism involving arsenic interstitials 
is the dominant process for the As-P and 
As-Sb interdiffusion in the material studied.\cite{Schultz1998} 

Similarly, annealing LT-GaAs $\delta$-doped 
with Sb was found to produce substantially greater 
compositional intermixing than annealing 
conventional stoichiometric GaAs 
similarly $\delta$-doped.\cite{Chaldyshev2001a}  
This enhancement of As-Sb interdiffusion was 
attributed to an oversaturation of arsenic 
interstitials in the LT GaAs 
samples, resulting from the balance of arsenic 
interstitials with arsenic clusters and all the 
other excess-arsenic-containing defects in the material.  
The effective activation energy 
for As-Sb interdiffusion in LT GaAs deduced from this work, 
0.6 $\pm$ 0.15 eV,\cite{Chaldyshev2001a} is 
reasonably close to the migration energy 
of 0.5 eV for arsenic interstitials deduced from 
annealing experiments on defects produced by 
electron irradiation,\cite{Bourgoin1988} 
as well as to migration energies subsequently 
ascribed to arsenic interstitial defects produced 
in GaAs by other means.  The concentration of arsenic 
interstitials required to produce 
sufficient oversaturation to eliminate completely any 
contribution of the interstitial formation 
energy to the activation energy for As-Sb intermixing 
measured in the LT GaAs sample was estimated 
to be roughly $10^{18}$~cm$^{-3}$, using Hurle's 
thermodynamic analysis in conjunction with the 
experimental data.\cite{Chaldyshev2001a} 

Theoretical attempts to obtain a picture of the 
microscopic structure and properties of the
lowest energy arsenic interstitial configuration(s) 
began with the work of Baraff and 
{Schl\"uter}, who used density functional Green's 
function calculations to investigate 
the energies of reactions creating native defects with 
$T_d$ symmetry in GaAs, including arsenic interstitials in the two 
tetrahedral sites.\cite{Baraff1985a} 
The effects of lattice relaxation were ignored.  
Baraff and {Schl\"uter} concluded that 
simple tetrahedral arsenic interstitials 
were less likely to occur than vacancy and
antisite defects under all equilibrium conditions, although 
they could not rule out the possibility 
that other, more complicated interstitial configurations 
might have a lower energy.\cite{Baraff1985a}

Jansen and Sankey calculated the formation 
energies for unrelaxed native defects 
with tetrahedral symmetry in GaAs, including 
arsenic interstitials in tetrahedral sites, 
using a density-functional pseudopotential 
method with a basis set of pseudo-atomic orbitals 
and a single special {\bf k}-point 
in supercells containing about 32 atoms.\cite{Jansen1989}  
In order to calculate
the formation energies for individual defects 
instead of reaction energies for defect 
reactions which conserve the number of atoms of each 
species, they were required
to choose a value for the arsenic chemical potential (or equivalently, 
for the gallium chemical potential).  An arbitrary value was chosen, 
corresponding to the condition that the formation energies for neutral
gallium vacancies and for neutral arsenic vacancies should be equal.  
Jansen and Sankey concluded that arsenic interstitials in tetrahedral sites
should be less numerous than vacancies and antisites in GaAs under 
equilibrium conditions,\cite{Jansen1989} 
in agreement with Baraff and {Schl\"uter}.
 
Zhang and Northrup used density functional theory (DFT) within the 
local density approximation (LDA) and supercells of about 
32 atoms to calculate the formation energies for vacancies, antisites, 
and tetrahedral interstitials in GaAs as a function 
of arsenic chemical potential, over the 
physically allowable range 
of chemical potentials, from Ga-rich to As-rich.\cite{Zhang1991} 
This physically allowable range is set by the heat of
formation of bulk GaAs and by the 
requirement that the arsenic chemical potential may not exceed the chemical 
potential of bulk arsenic, since the material is in equilibrium with arsenic 
precipitates in the arsenic-rich limit.  
The atomic coordinates were allowed to relax
in these calculations, within the constraints imposed by the tetrahedral 
symmetry.  In agreement with the previous work,
Zhang and Northrup found that antisites and/or vacancies should be
more numerous than arsenic interstitials in tetrahedral sites under 
all equilibrium conditions.\cite{Zhang1991} 

Chadi used DFT-LDA calculations and 33-atom supercells to investigate 
many different types of bonding configurations for self-interstitials 
in GaAs, including various split interstitials, as well as hexagonal, 
two-fold coordinated, and tetrahedral interstitials, all fully 
relaxed within the constraints of the chosen symmetry.\cite{Chadi1992}    
He found that the lowest energy configuration for arsenic interstitials
in the neutral or $-1$ charge state is a split interstitial consisting of 
two As atoms sharing an arsenic lattice site, displaced from this site
in opposite directions along a $<$110$>$-like axis, while the lowest energy 
configuration for positively charged arsenic interstitials
in the +1 or +2 charge state is a split interstitial consisting of 
an As atom and a Ga atom sharing an gallium lattice site, 
displaced from this site
in opposite directions along a $<$100$>$-like axis.  
Since we will be interested below primarily in arsenic interstitials
in semi-insulating or $n$-type GaAs, we will use the notation 
As$_i$-As for the interstitial with two atoms 
sharing an arsenic site and aligned along a $<$110$>$-like axis, 
which should be the lowest energy
interstitial configuration in semi-insulating or $n$-type material.

Chadi also showed that neutral arsenic interstitials, which 
have unpaired spins, are unstable relative to formation of a pair of 
+1 and $-1$ charged interstitials ---  
{\em i.e.}\  arsenic interstitials form a negative-$U$ system. 
This suggested that arsenic interstitials may 
not be observable in EPR experiments.\cite{Chadi1992}  Chadi reported the 
relative energies for the most energetically favorable arsenic
interstitial configurations in each of these charge states, including 
in each case a number of metastable configurations somewhat 
higher in energy than 
the lowest energy configurations, all
of which were more complicated than the simple tetrahedral 
configurations.\cite{Chadi1992}  However, since Chadi did not report 
absolute interstitial formation energies as a function of 
arsenic chemical potential, no comparison with the formation energies
of defects involving a different number of excess arsenic atoms, such 
as arsenic antisites, was possible from this work. 

Landman {\em et al.}\ investigated the relative formation energies of 
the point defects containing excess As, As$_{\mathrm{Ga}}$, 
$V_{\mathrm{Ga}}$, and the lowest energy As$_i$ configuration in 
semi-insulating or $n$-type, As-rich GaAs, As$_i$-As,\cite{Landman1997} 
using DFT-LDA-based calculations with the Harris-Foulkes functional 
and a basis of pseudo-atomic 
orbitals.\cite{Sankey1989, Tsai1992}  They placed the defects in 64-atom 
supercells, and estimated 
summations in {\bf k}-space by using a single Chadi and Cohen
special point.\cite{Chadi1973}  Since Harris-Foulkes, pseudoatomic-orbital 
calculations do not give as accurate results for semiconductor heats of 
formation or for the relative energies of compound semiconductor and 
pure, metallic phases as fully self-consistent DFT-LDA calculations
with a sufficiently large basis of plane waves, they did not use this 
method to calculate the arsenic
chemical potential in the arsenic-rich limit.
Instead, the relative formation energies of the tetrahedral As$_i$ with 
arsenic nearest neighbors, As$_{\mathrm{Ga}}$, and 
$V_{\mathrm{Ga}}$ in the arsenic-rich limit for the chemical potential 
were taken from information given by Zhang and 
Northrup,\cite{Zhang1991} and the relative formation 
energy of As$_i$-As was determined by Landman {\em et al.}'s 
result that the neutral As$_i$-As is 4 eV lower in 
energy than the unrelaxed, neutral 
tetrahedral As$_i$ with arsenic nearest neighbors.  
Since the tetrahedral interstitial was found to 
be unstable, relaxing to another configuration in 
Landman {\em et al.}'s calculation, 
they were obliged to compare their results for the ideal,
unrelaxed tetrahedral interstitial 
to Zhang and Northrup's results for a tetrahedral interstitial 
which had been relaxed 
while constrained to keep its tetrahedral symmetry.  
This led to an additional uncertainty 
in the relative formation energies between zero and 0.8 eV.\cite{Landman1997}  
However, Landman {\em et al.}\ concluded that the lowest 
energy split As$_i$ may have a concentration 
approaching that of As$_{\mathrm{Ga}}$ for certain Fermi 
levels.\cite{Landman1997} 

Since the theoretical investigations described above have been 
carried out over a long period 
of time, it has gradually become possible not only to include 
lattice relaxation 
and to investigate more complicated interstitial configurations, 
but also to do 
more accurate calculations, 
using larger unit cells and better sets of {\bf k}-points for the summations 
over {\bf k}-space.    
P\"oykko {\em et al.}\ showed how sensitive calculated defect 
properties can be to the 
{\bf k}-space sampling method and supercell size
in their investigation of the 
$V_{\mathrm{As}}$-Si$_{\mathrm{Ga}}$ complex in GaAs.\cite{Poykko1996b}  
They found that the use of the $\Gamma$ point can 
produce misleading results even when supercells are 64-atoms in size, 
reinforcing the conclusions of Makov that the $\Gamma$ point produces 
particularly slowly converging results with respect to cell 
size.\cite{Makov1996}
So it is essential to use a special point mesh in this type of 
calculation.
Furthermore, Puska {\em et al.}\ concluded that cell sizes of 128 to 216 atoms 
are needed to properly assess the physical properties of the silicon 
vacancy in bulk silicon, because of the dispersion of energy levels 
and long range ionic relationships.\cite{Puska1998}

In this paper, we investigate
the combined effects of cell size and {\bf k}-space sampling
on the formation energy, charge state transitions, atomic 
relaxations, and characterization of localized defect states 
for arsenic self-interstitials in GaAs. 
Because of the more ionic nature of the material and the complicated
split interstitial defect 
structure, comparison of these results for interstitials in GaAs to 
the previous results for vacancies in silicon\cite{Puska1998} can 
enhance our understanding of the range of behavior for different 
defects in different materials.
We compare the
formation energy of the lowest energy arsenic interstitial in $n$-type 
or semi-insulating GaAs, As$_i$-As, with the formation energies 
of As$_{\mathrm{Ga}}$ 
and $V_{\mathrm{Ga}}$ at the arsenic-rich end 
of the range of physically allowed chemical potentials, all calculated by 
state-of-the-art DFT 
pseudopotential\cite{Bockstedte1997} calculations, using the larger 
supercells and sets of special {\bf k}-points which we have determined 
to be necessary.  We conclude our study by discussing 
the relative concentrations of these defects in equilibrium 
in As-rich, $n$-type or semi-insulating GaAs at growth
temperatures,
and reporting the computed charge transition levels 
and expected electrical behavior of As$_i$-As as a function of Fermi level.

\section{Computational Method}
\label{sec:ComputationalMethod}
We have used the molecular dynamics code developed at the Fritz Haber 
Institut (FHIMD)\cite{Bockstedte1997}
for this investigation, using
density-functional theory (DFT) \cite{Hohenberg1964} within the 
local density approximation (LDA), with
the Ceperley-Alder \cite{Ceperley1980} form for the exchange and 
correlation potentials as parameterized by Perdew and 
Zunger.\cite{Perdew1981}
The core electrons are treated in the frozen-core approximation 
and the ion cores are replaced by fully-separable \cite{Kleinman1982} 
norm-conserving pseudopotentials.\cite{Hamann1989}  
Plane waves are included up to the energy cutoff of 10~Ry. 
The atoms are allowed to relax until the force 
components are are less than $5 \times 10^{-4}$~hartrees per 
bohr radius and the zero temperature formation energies change 
by less than $5 \times 10^{-6}$~hartrees per step for at 
least 100 steps.

To evaluate the defect formation energy, 
we used the formalism of Zhang and 
Northrup,\cite{Zhang1991} which gives for the formation energy
in the As-rich limit at zero temperature
\begin{eqnarray}
\Delta E_f = E(N_{\mathrm{Ga}}, N_{\mathrm{As}}, q) 
  - N_{\mathrm{Ga}} \mu_{\mathrm{GaAs}} \nonumber \\
  - (N_{\mathrm{As}} - N_{\mathrm{Ga}}) \mu_{\mathrm{As (bulk)}}
  + q \epsilon_F\, .
\label{eq:formation}
\end{eqnarray}
Here $q$ electrons have been 
transferred to a reservoir at the Fermi energy $\epsilon_F$ in order to 
produce a defect in the desired charge state. 
$E(N_{\mathrm{Ga}}, N_{\mathrm{As}}, q)$ is the zero-temperature total 
energy produced by the {\em ab initio} code for a supercell containing
the desired defect,
the chemical potential $\mu_{\mathrm{GaAs}}$ is the energy 
per atomic pair of bulk GaAs, and the arsenic chemical potential in 
the As-rich limit, $\mu_{\mathrm{As (bulk)}}$, is the energy per atom 
of pure bulk As computed using the same {\em ab initio} code and 
pseudopotentials.
$N_{\mathrm{Ga}}$ and $N_{\mathrm{As}}$ are the numbers of atoms of 
each species in the supercell containing the defect.
We will discuss the effect of temperature, which can be important for defect 
concentrations, in Section \ref{sec:ResultsAndDiscussion}.

Because the zero of the energy levels floats freely, \cite{Kleinman1981}
results from different DFT supercell calculations must be aligned in 
order to obtain the correct charge transition levels.  
We apply the procedure outlined by Kohan {\em et al.}\  
\cite{Kohan2000} in which we first compute the difference between the 
electrical potential in the supercell with the neutral defect and 
the electrical potential for the corresponding bulk crystal supercell, 
averaged over parallel planes, as a function 
of position along a line normal to the planes.
Far from the defect within the supercell, this difference becomes a 
constant.  In order to make the potential far from the neutral defect 
equal to the corresponding potential in the bulk cell, a 
uniform shift is applied to the potential and the energy levels, 
yielding the proper alignment of the energy levels of the defect with the 
energy levels of ideal bulk crystal supercell.   
The same shift is applied for all charge states of a given defect.

A well-known shortcoming of the LDA is that it underestimates the 
band gaps of materials.  
The typical method for dealing with this problem is to 
simply shift the conduction band states up uniformly by the amount 
needed to reproduce the experimental gap, using the so-called 
`scissors operator'.\cite{Baraff1984}  
More recently, an analytically-based model justifying the rigid
shifting upward of all conduction band states by a scissors-type correction
has been shown to produce good results for a large number
of semiconductors.\cite{Johnson1998a}
Since the LDA can produce similarly large errors in the energies of 
the deep defect states, it is also important to correct for these errors 
when determining where the charge transition levels corresponding to 
deep defect states lie in the experimental gap.  
Unfortunately, a full GW calculation,
\cite{Hedin1965a,Hedin1969a,Hybertsen1986a,Godby1988a,Rohlfing1993a}
which would correct these errors, 
is not currently possible for the large supercells needed for studies 
of defects.  Therefore, we apply the same upward shift to the defect 
states with predominantly conduction band character as is applied to 
the conduction band states themselves, while leaving the defect states 
with predominantly valence band character fixed relative to the valence 
band edge.

\section{Results and Discussion}
\label{sec:ResultsAndDiscussion}
\subsection{Defect formation energies, charge transition levels, 
and localized defect states}
\label{subsec:FormationEChargeTransitLevelsAndDefectStates}
We used cubic supercells with dimensions of both two and three times 
the computationally determined bulk lattice constant,
corresponding to bulk cells of 64 and 216 atoms,
along with three different Brillouin zone (BZ) sampling schemes
to examine the effects cell size and sampling scheme have
on the formation energies and transition levels for the As$_i$-As.  
In order to investigate the most efficient choice of {\bf k}-points to 
obtain good results, we have used a $1^3$ 
Monkhorst-Pack (MP) mesh,\cite{Monkhorst1976} a $2^3$ MP mesh, 
and the $\Gamma$ and $L$ points, which was recommended as a good minimal 
set of {\bf k}-points for cubic supercells with no particular 
symmetry within the supercell,\cite{Makov1996} and which has subsequently 
been used in defect calculations, {\em e.g.}\  to study the structures 
associated with dopants in highly $n$-doped Si.\cite{Chadi1997}
Calculations comparing different summation schemes for vacancies in Si 
show that use of the $\Gamma + L$ points produces a reasonably 
well-converged formation energy in a 64-atom supercell.\cite{Puska1998}

This earlier work on the vacancy in silicon has shown that the 
neutral vacancy formation energy computed with different
{\bf k}-space sampling schemes converges at different rates with 
respect to supercell size.\cite{Puska1998}
However, an acceptably converged value for the neutral vacancy formation 
energy can be attained more easily than an acceptably converged 
description of the 
charge transition levels and the atomic relaxations and defect 
symmetry for different charge states.
\cite{Puska1998}

\begin{table}
\caption{Formation energies for excess-arsenic-containing defects 
computed in the 
As-rich limit.  These were calculated with supercells corresponding
to the bulk 216 atom cell.  The values in the last column are computed 
with the Fermi level pinned at the calculated (+1/0) transition 
level of the As 
antisite defect.}
\label{table:formationenergies}
\begin{ruledtabular}
\begin{tabular}{ccccc}
Defect& {\em k}-space&{Charge}& {Formation}& {Formation}\\
& sum& {state}& {energy (eV)}& {energy (eV)}\\
& & & {$\epsilon_F$ at VBM}& {semi-insulating $\epsilon_F$}\\ 
\hline
As$_{\mathrm{Ga}}$ & MP $2^3$ & $+2$ & 0.9 & 2.0\\
As$_{\mathrm{Ga}}$ & MP $2^3$ & $+1$ & 1.3 & 1.8\\
As$_{\mathrm{Ga}}$ & MP $2^3$ & $ 0$ & 1.8 & 1.8\\
\hline
As$_{\mathrm{Ga}}$ & MP $1^3$ & $+2$ & 0.9 & 1.8\\
As$_{\mathrm{Ga}}$ & MP $1^3$ & $+1$ & 1.1 & 1.6\\
As$_{\mathrm{Ga}}$ & MP $1^3$ & $ 0$ & 1.6 & 1.6\\
\hline
As$_{i1}$ & MP $2^3$ & $ 0$ & 6.9 & 6.9\\
\hline
As$_{i2}$ & MP $2^3$ & $ 0$ & 6.2 & 6.2\\
\hline
As$_i$-As & MP $2^3$ & $+1$ & 3.5 & 4.1\\
As$_i$-As & MP $2^3$ & $ 0$ & 3.8 & 3.8\\
As$_i$-As & MP $2^3$ & $-1$ & 4.4 & 3.8\\
\hline                              
As$_i$-As & MP $1^3$ & $+1$ & 3.4 & 3.8\\ 
As$_i$-As & MP $1^3$ & $ 0$ & 3.7 & 3.7\\
As$_i$-As & MP $1^3$ & $-1$ & 4.1 & 3.7\\
\hline
As$_i$-As & $\Gamma + L$ & $+1$ & 3.6\\
As$_i$-As & $\Gamma + L$ & $ 0$ & 3.6\\
As$_i$-As & $\Gamma + L$ & $-1$ & 4.1\\
\hline
$V_{\mathrm{Ga}}$ & MP $2^3$ & $ 0$ & 2.9 & 2.9\\
$V_{\mathrm{Ga}}$ & MP $2^3$ & $-1$ & 3.0 & 2.5\\
$V_{\mathrm{Ga}}$ & MP $2^3$ & $-2$ & 3.1 & 2.1\\
$V_{\mathrm{Ga}}$ & MP $2^3$ & $-3$ & 3.4 & 1.8\\
\end{tabular}
\end{ruledtabular}
\end{table}

To augment our understanding of the effects of cell size and 
sampling scheme, we computed the formation energies of the 
different charge states
for the fully relaxed split interstitial As$_i$-As in GaAs.
Charge was balanced by a uniform background to avoid long range 
Coulomb interactions between the supercells.

Table~\ref{table:formationenergies} lists the formation energies 
we obtained for all the excess-arsenic-containing elementary point 
defects in the arsenic-rich limit, with GaAs in equilibrium with 
bulk arsenic, including complete results for the various 
{\bf k}-space sums for As$_i$-As.
For comparison, the formation energies of the unrelaxed, 
ideal tetrahedral As interstitials with As neighbors (As$_{i1}$) and 
with Ga neighbors (As$_{i2}$) are also shown.  These tetrahedral 
interstitials are unstable, and will relax to other configurations 
if allowed to break their tetrahedral symmetry.

Table~\ref{table:formationenergies} displays 
formation energies evaluated for the Fermi level at the valence band 
maximum (VBM), and also for the Fermi level pinned at the 
calculated position of the 
(+1/0) charge state transition of the As$_{\mathrm{Ga}}$, which is at  
$\mathrm{VBM} + 0.54~ \mathrm{eV}$ for the $2^3$ MP mesh and 
$\mathrm{VBM} + 0.45~ \mathrm{eV}$ for the $1^3$ MP mesh.  This 
choice of Fermi level was based on the experimental finding that 
there can be high concentrations of As$_{\mathrm{Ga}}$ in As-rich 
GaAs, and that these high concentrations of As$_{\mathrm{Ga}}$ 
can pin the Fermi energy near this transition level.  
Using either choice of Fermi level as a reference, the formation 
energies and equilibrium concentrations of the defects can be 
determined as a function of Fermi energy (or doping level).  

\begin{figure}
\includegraphics{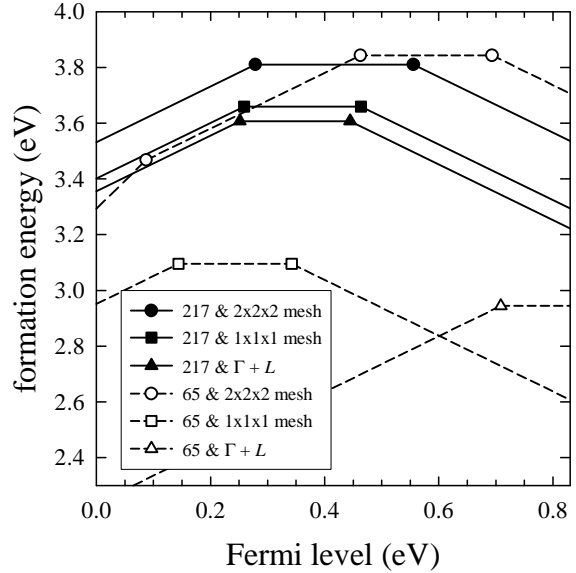}
\caption{Defect formation energies for the As split interstitial
computed in the As-rich limit.  Comparisons are presented for different
{\bf k}-space sums and two supercell sizes.
Dashed lines are used
for cells containing 65 atoms, and solid lines for 217 atom cells.
Transition levels between the different charge states are marked 
with circles for the $2^3$ MP mesh, 
squares for the $1^3$ MP mesh, and triangles for the $\Gamma + L$ 
{\bf k}-space sum.}
\label{fig:formationenergymethods}
\end{figure}

In Fig.~\ref{fig:formationenergymethods}, we present the results of 
the formation energy calculations for As$_i$-As in which both 
cell size and {\bf k}-space 
sampling methods have been varied.  This figure shows the formation 
energy for the As$_i$-As in its preferred charge state.  For Fermi 
levels in the range where the neutral As$_i$-As is preferred, 
the formation energy is independent of Fermi level.  If the Fermi 
level is decreased past the (0/$+1$) charge transition level, so 
that the +1 charge state is preferred, the formation energy vs. 
Fermi level curve has a slope of +1, as can be seen from 
Eq.~(\ref{eq:formation}), and if the Fermi 
level is increased past the (0/$-1$) charge transition level, so 
that the $-1$ charge state is preferred, the formation energy vs. 
Fermi level curve has a slope of $-1$.  

The most obvious feature in Fig.~\ref{fig:formationenergymethods} 
is the wider 
variation in the energies computed using the smaller supercell.  
Convergence with cell size is also visibly slower when using the 
$1^3$ MP mesh or the 
$\Gamma + L$ points, which use a less finely spaced set of 
{\bf k}-points than the $2^3$ MP mesh to cover the Brillouin zone.
It is clear that the 217 atom cell with the $2^3$ MP mesh is well 
converged.  We see that the two less finely spaced sampling 
schemes produce 
somewhat converged results in the 217 atom cell, as does the $2^3$ MP 
mesh in the 65 atom cell.
In agreement with previous results for the vacancy in bulk 
Si,\cite{Puska1998} we find 
that different sampling schemes can be used to produce either an 
attraction between defects ({\em i.e.}\  a lowering of energy 
in the smaller supercell), as in the $1^3$ MP or the $\Gamma + L$ cases, 
or a charge-state dependent repulsion or attraction between defects, 
as in the $2^3$ MP case.

When comparing the results for the 65-atom and the 217-atom cells, 
using the $2^3$ MP mesh, we see that the charge transition levels 
converge less rapidly than the neutral defect formation energy 
with increasing cell size, due to long-range interactions of the 
electrons in the localized defect state with the charged defects 
in neighboring cells.  The neutral As$_i$-As formation energy 
changes by less than 0.1 eV when the cell size is reduced from 
217 to 65 atoms, while the charge transition levels move by about 
0.4 eV, when using the $2^3$ MP mesh.  This trend is not seen in 
the less-well-converged results obtained using the $1^3$ MP mesh 
or the $\Gamma + L$ points, where the neutral As$_i$-As 
formation energies move by 0.5 to 0.6 eV when the cell size is 
reduced from 217 to 65 atoms, perhaps due to interactions of the 
deep defect level with the band edges.

\begin{figure}
\includegraphics{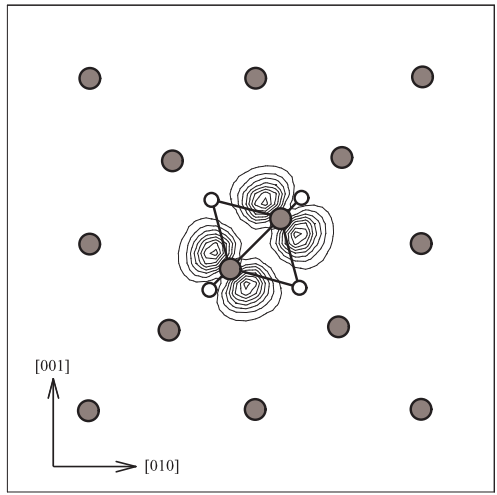}
\caption{This contour plot shows levels of constant charge density 
for the localized defect state of the neutral As$_i$-As 
in the 217 atom supercell, evaluated in 
the plane containing the two defect As atoms, which is 0.48\AA\ above 
the As lattice site associated with the defect.
The dark circles represent the positions of the As atoms, including the 
defect atoms and the As atoms of the original lattice plane.  
The light circles show the locations of the neighboring Ga atoms
projected onto the plane.}
\label{fig:pibond}
\end{figure}

In the neutral charge state, the topmost filled electronic level is 
half filled.  This level corresponds to a localized state centered on
the split interstitial oriented along a $<$110$>$-like direction.  
In Fig.~\ref{fig:pibond}, the charge density associated with this 
defect state is shown in a plane parallel to an arsenic lattice 
plane, but slightly above this plane, so that it includes the 
two arsenic atoms of the split interstitial, which have relaxed 
slightly away from the ideal lattice site.  This plot clearly shows 
that the 
defect state is localized in $p$-like orbitals which appear to be 
forming a $\pi$ antibonding state, as evident from the node in the 
charge density midway between the two As atoms forming the split 
interstitial.  
This result is corroborated by an examination of the 
characterization of the state, using the 217-atom cell and the 
$2^3$ MP mesh, which shows that  in contrast to the 
very extended character of the bulk-like states, 20\% of the 
defect state is localized in the $p_y$ and $p_z$ orbitals of the 
two As atoms of the neutral defect.  Since the charge density 
in this defect state points roughly along the direction of the 
As-Ga bonds between the defect As atoms and the two Ga atoms bonded
to both of these As atoms,
we note that this state makes a bonding contribution to these As-Ga
interactions as well as making an antibonding contribution to 
the interaction between the As atoms of the defect.

The characterization of the deep defect state in the 217-atom 
supercell is not very sensitive to the {\bf k}-space sampling 
method used.  For example, 20\% of the state is also found to be 
localized in the $p_y$ and $p_z$ orbitals of the 
two As atoms of the neutral defect when the $\Gamma + L$ sampling is 
used.

In smaller supercells, the defect state  
interacts with its images and becomes less localized.  One way to 
observe this is to examine the variation in energy of the defect 
state across {\bf k}-space, {\em i.e.}\  the dispersion of the state.  
We characterize this variation by computing the difference between 
the highest and lowest energy obtained from a {\bf k}-space survey 
along the $\Delta$, $\Sigma$, $\Lambda$, and $T$ lines in the cubic 
Brillouin zone.
The dispersion measured in this way is 0.1~eV in the 217 atom supercell, 
and 0.5~eV in the 65 atom cell.
This interaction of the defects in neighboring supercells contributes 
to the variation in the positions of the charge 
transition levels between the different cell sizes and sampling 
schemes, noted above, and supports the conclusion that the larger 217 
atom supercell should be used for accuracy in describing the charge 
transition levels and deep defect states.

\subsection{Defect atomic structure and relaxation}
\label{subsec:AtomicStructureAndRelaxation}
\begin{figure}
\includegraphics{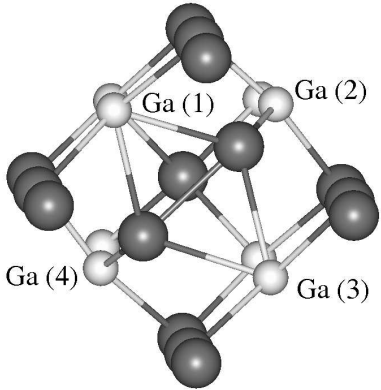}
\caption{The neutral As$_i$-As defect from the 217-atom 
supercell is shown (viewed from a direction slightly displaced from the 
[100] direction) with larger dark spheres representing the As atoms and 
smaller light spheres representing the Ga atoms.}
\label{fig:splitint}
\end{figure}

The detailed structure of the atomic positions is also expected to 
be dependent on the cell size and {\bf k}-space sampling approach.  
However, in contrast to the behavior observed for the vacancy in 
silicon,\cite{Puska1998} where the defect symmetry and atomic 
relaxations are very sensitive to the supercell size and {\bf k}-space 
sampling, we find that the atomic structure of the As$_i$-As defect is 
remarkably similar for the 65-atom and 217-atom supercells, and also 
depends little on the {\bf k}-space sampling approach.  
In Fig.~\ref{fig:splitint} we show the structure of the neutral 
split interstitial oriented along the $<$011$>$ direction,
viewed from the [100] direction.  The two Ga atoms 
labeled Ga(1) and Ga(3) are bonded to both As atoms of the 
defect, while those labeled Ga(2) and Ga(4) are bonded to only 
one of the defect atoms.
The structure exhibits $C_{2v}$ symmetry in all charge states and 
returns to this symmetry when the atomic coordinates are perturbed 
from their equilibrium positions and allowed to relax.

Although the local lattice expansion introduced by the additional atom 
of the interstitial might be expected to converge slowly with 
supercell size, we find that reasonable convergence is more easily 
reached for the atomic positions than for the electronic properties 
discussed in the previous section.
The As-As and As-Ga distances found in the 65 atom supercell 
differ by less than 0.01~\AA\  from those found in the 217 atom supercell, 
using the 2$^3$ MP mesh.
The distance between the two As atoms of the As$_i$-As defect is 2.39~\AA.
(For comparison, the GaAs bulk nearest neighbor distance in this 
calculation is 2.41~\AA.)
The distance between Ga(1) or Ga(3) and either of the As atoms 
of the defect is  
2.60~\AA.  The distance between Ga(2) or Ga(4) and the nearest As of 
the split interstitial is 2.32~\AA.

The two different supercell sizes produce slightly different 
results in the position of the center of mass of the two As atoms 
of the split interstitial.  
The center of mass of the pair of As atoms is shifted away from the 
bulk lattice site by 0.48~\AA\  in the [100] direction toward the 
plane containing Ga(1) and Ga(3) in the case of the 217 atom 
supercell, using the 2$^3$ MP mesh.
This shift is 0.50~\AA\  in the smaller supercell.  Since the nearest 
neighbor distances are very similar in the two supercells, this 
difference is accomplished through variation of bond angles.  The angle 
between the bonds from one of the As atoms to the Ga(1) and Ga(3) 
atoms is $107.7^\circ$ in 
the small supercell and $106.8^\circ$ in the large supercell.

We find that the atomic structure of the 
As$_i$-As defect depends 
very little on the {\bf k}-space summation method, as can be shown by 
examining the bond lengths between the As atoms of the defect 
and between those As atoms and the four neighboring Ga atoms
in the 217 atom supercell.  For the neutral As$_i$-As defect, these bond 
lengths change by less than 0.1\% when the summation method is changed. 

For the $+1$ charge state of the defect,
we find a small but perceptible dependence of the bond lengths on 
the {\bf k}-space 
summation method.
The As-As distance obtained using the $2^3$ MP mesh is 
about 1\% smaller than that obtained using the $\Gamma + L$ sampling.
The bond between either of 
the As atoms and the Ga(1) or Ga(3) atom
is about 1\% longer when using the $2^3$ MP mesh than when using 
the $\Gamma + L$ summation method.  
The distance between the Ga(2) or Ga(4) atom and the nearest As atom 
of the split interstitial is about 0.5\% longer for 
the $2^3$ MP mesh calculation than for the $\Gamma + L$ calculation.

There is a similar but weak effect (under 0.5\%) 
in the defect bond lengths observed in the $-1$ state, with the roles of 
the $2^3$ MP mesh and the $\Gamma + L$ points reversed --- {\em i.e.}\ 
the $2^3$ MP mesh now gives a larger As-As distance and smaller As-Ga 
distances.
The $1^3$ MP mesh produces results between those of the other two 
{\bf k}-space summation methods.  

The effect of these small variations in bond lengths 
is to reduce the changes in bond length  
seen in the $2^3$ MP mesh calculations when the defect becomes charged, 
if one of the other {\bf k}-space sampling methods is used instead.
In particular, the $\Gamma + L$ sampling is observed to produce a somewhat
smaller dependence of these bond lengths on charge state.  This dependence 
of bond lengths on the charge state of the defect is seen below to 
result from the changes in occupation of the deep defect state when the 
charge state is changed, and from the bonding or antibonding character 
of this state for particular bonds.    

Focusing on results for the 
217 atom cell with the $2^3$ MP 
mesh, we observe that the distance between the atoms of the defect 
has a significant dependence on the charge state of the system.  
The As-As bond expands 
to 2.47~\AA\  (about 3.5\% compared to the neutral state bond length),
when the system is allowed to relax in the $-1$ charge state.  
This can be easily understood, since the antibonding defect state on the 
two As atoms is doubly occupied and can cause a stronger repulsive 
contribution to the interaction between these two atoms for the $-$1 
charge state of the defect, while it is only singly occupied for the 
neutral state of the defect.  

When the system is allowed to relax in the $+1$ state 
the As-As bond shrinks to 2.31~\AA, a contraction of about 3\%.  
This can also be easily understood, since the 
antibonding defect state on the two As atoms of the defect is completely 
unoccupied for the +1 charge state, so it no longer makes 
a repulsive contribution to the interaction between the two As atoms.  
The bond length between Ga(1) or Ga(3) and either As atom
of the defect 
is 3.5\% longer and the bond length between Ga(2) or 
Ga(4) and the nearest defect As 
atom is 1.9\% longer in the $+1$ state than in the neutral state.  
Since the deep defect state acts as a bonding state for the 
As-Ga bonds between the defect As atoms and Ga(1) and Ga(3), 
as discussed above in 
Section~\ref{subsec:FormationEChargeTransitLevelsAndDefectStates},
it is easy to understand why these 
bonds are longer in the $+1$ state, when the defect state is 
fully unoccupied and can no longer contribute to the strength of 
these bonds. 
 
We note that because of the contribution of the defect state to the 
As-Ga bonds between the defect As atoms and Ga(1) and Ga(3), these 
two Ga atoms move 
significantly when the charge state is changed, changing the 
occupation of the defect state.  These Ga atoms are 
about 4\% closer to each other in the 
$-1$ state and about 5.5\% farther apart in the $+1$ state than 
in the neutral state.  The other two 
Ga atoms, each bonded to only one defect As atom, do not move in 
response to the change in charge state.  

In performing these calculations, we fixed the lattice constant at the 
value determined through minimization of the energy of the bulk crystal.  
While the ideal calculation should include a full lattice constant 
determination with each change of defect configuration, for simplicity 
we did not perform this relaxation.  This may be 
deemed a reasonable choice in light of evidence presented by Puska, 
{\em et al.}\cite{Puska1998}\  for {\em ab initio} 
supercell calculations in the LDA, 
using supercells of sizes comparable to ours, in which vacancies in 
bulk Si are found to alter the lattice constant by around 0.2\%, while 
artificially introduced distortions in the lattice constant of up 
to 1\% are seen not to affect their reported results significantly.  

\subsection{Relative defect concentrations in equilibrium}
\begin{figure}
\includegraphics{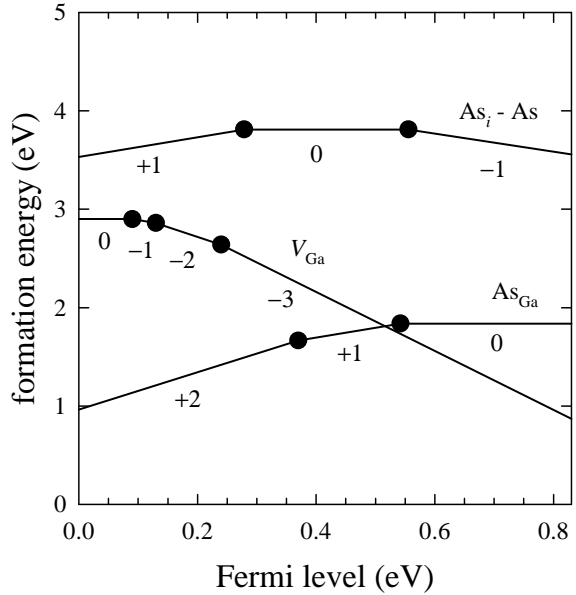}
\caption{Defect formation energies for selected defects over the 
calculated band gap in the As-rich limit.  Zero Fermi level 
corresponds to the valence band maximum.}
\label{fig:formationenergies}
\end{figure}

We now compare our well-converged results for the formation energies 
of the elementary excess-arsenic-containing point defects, As$_{\mathrm{Ga}}$, 
$V_{\mathrm{Ga}}$, and As$_i$-As (the most favorable As$_i$ 
configuration in semi-insulating or $n$-type GaAs), computed using 
the large supercell and the $2^3$ MP mesh.  These formation energies 
in the As-rich limit, corresponding to GaAs in equilibrium with 
bulk arsenic, are presented as a function of Fermi energy in 
Fig.~\ref{fig:formationenergies}.  The formation energies for two 
specific choices of Fermi energy have also been listed in 
Table~\ref{table:formationenergies} in 
Section~\ref{subsec:FormationEChargeTransitLevelsAndDefectStates}.

In Fig.~\ref{fig:formationenergies}, 
we can see that the $V_{\mathrm{Ga}}$ and 
As$_{\mathrm{Ga}}$ defects possess significantly lower formation 
energies than the As$_i$-As for all Fermi energies.  For example, 
the formation 
energy for As$_{\mathrm{Ga}}$ is seen to be at least 2 eV lower 
than the formation 
energy for As$_i$-As for all Fermi energies.  Small 
uncertainties in the formation energy should not alter this 
strong qualitative ordering of the formation energies or the 
prediction 
based on this ordering that equilibrium 
concentrations of As$_i$-As should be significantly lower than 
equilibrium concentrations of As$_{\mathrm{Ga}}$, as discussed 
below.

To estimate equilibrium concentrations of the 
excess-arsenic-containing defects we 
begin with the usual expression 
\begin{equation}
C = Ne^{-\Delta E_f/k_BT}e^{S_f/k_B}e^{-P\Delta V_f/k_BT}\, ,
\label{eq:concentration}
\end{equation}
where $N$ is the number of sites at which the defect can form in the 
crystal per unit volume, 
$\Delta E_f$ is the total energy of formation of the defect, 
$k_B$ is the Boltzmann constant, and
$T$ is the temperature. 
The formation entropy of the defect is $S_f$, $P$ is the pressure, and 
$\Delta V_f$ is the change in the crystal volume associated with the 
defect formation.

We note that the defect formation energy $\Delta E_f$ for charged 
defects, as given by the formula in Eq.~(\ref{eq:formation}), has an 
explicit dependence on the Fermi energy, in addition to its dependence 
on the calculated energies for defect formation at zero Fermi energy.  
Therefore we must compute the Fermi energy self-consistently, in 
order to determine the native defect concentrations present in a 
particular sample.  If any electrically active 
impurities or dopants are present in the material, the concentrations 
of these impurities or dopants in all charge states must also be 
taken into account.  We must set up the charge balance equation, 
requiring that the free electron and hole concentrations (which also 
depend on the Fermi level) must cancel out any net charge resulting from 
the concentrations of all positively and negatively charged defects and 
impurities.  This equation can then be solved to determine the Fermi 
level.  Once the Fermi level is known, it may be used to determine 
the formation energies and the resulting equilibrium concentrations of 
all the defects present.

If the defect formation energy $E_f$ for the most energetically 
favorable charge state of As$_i$-As in a particular sample is 
within $k_BT$ of the formation energy of the most favorable charge state of 
As$_{\mathrm{Ga}}$, we may expect that the equilibrium concentrations 
of these two defects are comparable, assuming that the effects of the 
entropy of formation $S_f$ and the change in the crystal volume 
$\Delta V_f$ associated with the defect formation can be neglected.  
We will now concentrate on the relative defect concentrations at 1500 K 
(near the melting point of GaAs), since defects with a higher formation 
energy such as As$_i$-As have their greatest chance to attain equilibrium 
concentrations comparable to those of more energetically favorable 
defects at high temperature.  We will estimate the effective corrections 
to the formation energy which occur at this temperature due to the entropy 
of formation and change in volume associated with the defects. 

First, to estimate the effect of the change in volume, we let $P$ be 
atmospheric pressure and overestimate $\Delta V_f$ to be the volume 
per bulk atom in the cell, which gives an effective correction to 
the defect formation energy $P\Delta V_f$ of $9\times 10^{-5}$~eV.  
We may safely neglect this correction.

Previous calculations\cite{Blochl1993} on defects in Si
found that the formation entropy $S_f$ is dominated by 
vibrational contributions, and that the formation 
entropies are $6 k_B$ and $5 k_B$ for the self-interstitial and the 
vacancy, respectively.
We note that the self-interstitial in silicon is a $<$110$>$ split 
interstitial with the same basic structure as As$_i$-As in GaAs.  
Therefore, in analogy to the results for defects in silicon,\cite{Blochl1993} 
we may assume that it is unlikely for the split interstitial As$_i$-As 
to have a very different formation entropy when compared to defects such 
as As$_{\mathrm{Ga}}$, which only contain atoms occupying lattice sites.  
If we let $S_f = 10 k_B$ (an overestimate) for As$_i$-As,
this gives rise to an effective reduction of the defect formation
energy by $S_f T$, or 1.3~eV at 1500~K.  
Even if we apply no reduction to the As$_{\mathrm{Ga}}$ formation energy 
due to entropy, this still leaves the effective formation energy 
about 0.7~eV  higher for As$_i$-As than for As$_{\mathrm{Ga}}$, 
producing equilibrium concentrations of 
As$_i$-As which are about 0.4\% those of As$_{\mathrm{Ga}}$ at 1500~K.  

We conclude that even using this extremely liberal estimate for the 
formation entropy of As$_i$-As and ignoring the formation entropy of 
As$_{\mathrm{Ga}}$ cannot lead to an As$_i$-As concentration 
approaching that of the 
antisites in thermal equilibrium.

\subsection{Defect electrical behavior}
Although the placement of the calculated charge transition levels 
in the experimental gap has an uncertainty far exceeding 0.1~eV due to 
the shortcomings of the LDA in calculating the gap, as discussed in 
Section~\ref{sec:ComputationalMethod}, our as-calculated 
band gap and charge transition levels are reported here to
0.1~eV (or 0.01~eV, for the closely spaced $V_{\mathrm{Ga}}$ levels), 
for the convenience of the reader who prefers not to 
read them off the picture in Fig.~\ref{fig:formationenergies}. 
For the As$_{\mathrm{Ga}}$, the (+2/+1) transition level appears at 
0.4~eV above the VBM, and the (+1/0) level is at 
$E_{\mathrm{VBM}} + 0.5~\mathrm{eV}$.  
For the As$_i$-As, the (+1/0) transition is at 
$E_{\mathrm{VBM}} + 0.3 \mathrm{eV}$, and the (0/$-1$) transition is at
$E_{\mathrm{VBM}} + 0.5 \mathrm{eV}$.
The levels for the $V_{\mathrm{Ga}}$ defect are at 
0.09~eV, 0.13~eV, and 0.2~eV above the VBM for the (0/$-1$), 
($-1$/$-2$), and ($-2$/$-3$) transitions, respectively.    
The calculated band gap of 0.8 eV is underestimated by 
0.7~eV compared to the experimental zero-temperature gap of 1.5~eV.

As discussed previously in Section~\ref{sec:ComputationalMethod}, we can 
get a rough estimate of where the charge transition levels fall 
within the experimental gap by shifting  the conduction band 
derived states (including the deep defect states with primarily 
conduction band character) by the amount needed to correct the gap, while 
leaving the defect states with predominantly valence band character 
fixed relative to the valence band edge.  Since the acceptor levels of 
the $V_{\mathrm{Ga}}$ are derived from the dangling bonds on the arsenic 
neighbors of the vacancy, which require three extra electrons to 
fill them, they should have the predominantly valence band character of 
anion dangling bond states.  In 
Section~\ref{subsec:FormationEChargeTransitLevelsAndDefectStates},
the deep defect state of the As$_i$-As was shown to have predominantly 
arsenic $p$-type character, similar to the character of the valence band 
edge states.  However, the As$_{\mathrm{Ga}}$ double donor defect 
state derives from an antibonding state of predominantly conduction band 
character, which has been lowered in energy due to the replacement of the 
original anion-cation bonds of the ideal crystal by anion-anion bonds 
between the antisite and its nearest neighbors.  This donor state is 
occupied by the two extra electrons contributed by the arsenic atom that 
has been substituted for a gallium atom, which cannot be accommodated in 
the bonding states of the valence band.  

\begin{figure}
\includegraphics{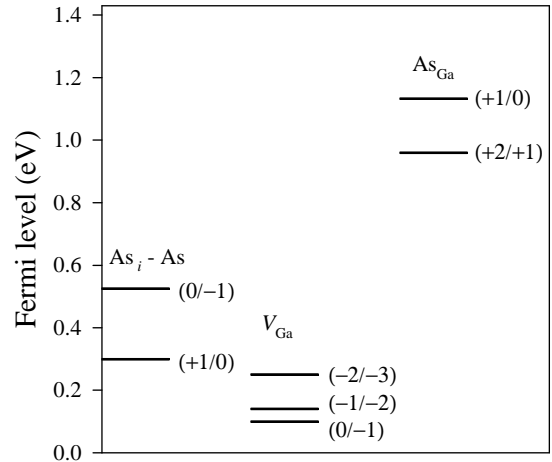}
\caption{Transition levels computed 
with a rigid shift applied to the conduction band and conduction band 
derived states to correct the LDA band gap underestimate.}
\label{fig:scissors1}
\end{figure}

We conclude that the charge transition levels of the V$_{\mathrm{Ga}}$ 
and the As$_i$-As should remain fixed relative to the valence band edge, 
while the donor levels of As$_{\mathrm{Ga}}$, which possess 
a conduction band character, should be shifted up together with the 
conduction band states.   In Fig.~\ref{fig:scissors1}, we show the charge 
transition levels for these defects corrected by the above procedure, 
using the room temperature gap of 1.4~eV.  The transition levels of 
As$_{\mathrm{Ga}}$ are shifted to 1.0~eV and 1.1~eV, in fortuitously
good agreement with the MCDA results identified with this 
defect in LT GaAs,\cite{Liu1995,Sun1992} 
although the transitions are both about 0.4~eV higher than those 
associated with As$_{\mathrm{Ga}}$ in melt-grown GaAs.\cite{Weber1982a} 

\section{Summary}

In this work we have shown that larger supercells and a better 
{\bf k}-space sampling than have been used in a large number of 
previous DFT defect 
calculations are required to give accurate 
results for the formation energies, charge 
transition levels, defect state 
properties, and atomic structure and relaxation for the arsenic split 
interstitial in GaAs.  In particular, we find that 217-atom 
supercells 
are necessary to get good results for the charge transition levels and 
the dispersion of the deep defect state, even 
though the arrangement of atoms in the structure is well converged in 
a 65-atom cell, particularly if one uses the finely spaced $2^3$ 
MP {\bf k}-space mesh.  

We have calculated formation energies for 
As split interstitials, Ga vacancies, and As antisites in As-rich 
GaAs using the larger supercells and better {\bf k}-point sampling which 
we have determined to be necessary.  Using these results, we find that the 
equilibrium
concentrations of arsenic interstitials will be substantially lower than 
equilibrium concentrations of arsenic antisites in As-rich, $n$-type or 
semi-insulating GaAs.
                  
\begin{acknowledgments}
This work was supported in part by AFOSR Grants
No.\ F49620-96-1-0167 and No.\ F49620-97-1-0472, and by 
grants of time on the Cray T3e computers at the 
DOD HPC Centers at NAVO and ERDC and 
at the NPACI San Diego and University of Texas Supercomputing Centers,
and on the NPACI AMD cluster at the University of Michigan.
We also thank G. Schwartz and M. Bockstedte for highly useful 
conversations.
\end{acknowledgments}


\begin{thebibliography}{50}
\expandafter\ifx\csname natexlab\endcsname\relax\def\natexlab#1{#1}\fi
\expandafter\ifx\csname bibnamefont\endcsname\relax
  \def\bibnamefont#1{#1}\fi
\expandafter\ifx\csname bibfnamefont\endcsname\relax
  \def\bibfnamefont#1{#1}\fi
\expandafter\ifx\csname citenamefont\endcsname\relax
  \def\citenamefont#1{#1}\fi
\expandafter\ifx\csname url\endcsname\relax
  \def\url#1{\texttt{#1}}\fi
\expandafter\ifx\csname urlprefix\endcsname\relax\def\urlprefix{URL }\fi
\providecommand{\bibinfo}[2]{#2}
\providecommand{\eprint}[2][]{\url{#2}}

\bibitem[{\citenamefont{Bourgoin et~al.}(1988)\citenamefont{Bourgoin, von
  Bardeleben, and Stievenard}}]{Bourgoin1988}
\bibinfo{author}{\bibfnamefont{J.~C.} \bibnamefont{Bourgoin}},
  \bibinfo{author}{\bibfnamefont{H.~J.} \bibnamefont{von Bardeleben}},
  \bibnamefont{and}
  \bibinfo{author}{\bibfnamefont{D.}~\bibnamefont{Stievenard}},
  \bibinfo{journal}{J. Appl.\ Phys.} \textbf{\bibinfo{volume}{64}},
  \bibinfo{pages}{R65} (\bibinfo{year}{1988}).

\bibitem[{\citenamefont{Hurle}(1999)}]{Hurle1999}
\bibinfo{author}{\bibfnamefont{D.~T.~J.} \bibnamefont{Hurle}},
  \bibinfo{journal}{J. Appl.\ Phys.} \textbf{\bibinfo{volume}{85}},
  \bibinfo{pages}{6957} (\bibinfo{year}{1999}).

\bibitem[{\citenamefont{Hurle}(1995)}]{Hurle1995a}
\bibinfo{author}{\bibfnamefont{D.~T.~J.} \bibnamefont{Hurle}},
  \bibinfo{journal}{Materials Science Forum}
  \textbf{\bibinfo{volume}{196-201}}, \bibinfo{pages}{179}
  (\bibinfo{year}{1995}).

\bibitem[{\citenamefont{Hurle}(1992)}]{Hurle1992}
\bibinfo{author}{\bibfnamefont{D.~T.~J.} \bibnamefont{Hurle}},
  \emph{\bibinfo{title}{Non-Stoichiometry in Semiconductors}}
  (\bibinfo{publisher}{Elsevier}, \bibinfo{address}{New York},
  \bibinfo{year}{1992}), p.~\bibinfo{pages}{47}.

\bibitem[{\citenamefont{Oda}(1992)}]{Oda1992}
\bibinfo{author}{\bibfnamefont{O.}~\bibnamefont{Oda}},
  \bibinfo{journal}{Semiconductor Science and Technology}
  \textbf{\bibinfo{volume}{7}}, \bibinfo{pages}{A215} (\bibinfo{year}{1992}).

\bibitem[{\citenamefont{Bublik}(1973)}]{Bublik1973}
\bibinfo{author}{\bibfnamefont{V.~T.} \bibnamefont{Bublik}},
  \bibinfo{journal}{Sov.\ Phys.\ Crystallog.} \textbf{\bibinfo{volume}{18}},
  \bibinfo{pages}{218} (\bibinfo{year}{1973}).

\bibitem[{\citenamefont{Charniy and Bublik}(1994)}]{Charniy1994a}
\bibinfo{author}{\bibfnamefont{L.}~\bibnamefont{Charniy}} \bibnamefont{and}
  \bibinfo{author}{\bibfnamefont{V.}~\bibnamefont{Bublik}},
  \bibinfo{journal}{J. Crystal Growth} \textbf{\bibinfo{volume}{135}},
  \bibinfo{pages}{302} (\bibinfo{year}{1994}).

\bibitem[{\citenamefont{Charniy
  et~al.}(1992{\natexlab{a}})\citenamefont{Charniy, Morozov, Bublik,
  Scherbachev, Stepantsova, and Kaganer}}]{Charniy1992a}
\bibinfo{author}{\bibfnamefont{L.~A.} \bibnamefont{Charniy}},
  \bibinfo{author}{\bibfnamefont{A.~N.} \bibnamefont{Morozov}},
  \bibinfo{author}{\bibfnamefont{V.~T.} \bibnamefont{Bublik}},
  \bibinfo{author}{\bibfnamefont{K.~D.} \bibnamefont{Scherbachev}},
  \bibinfo{author}{\bibfnamefont{I.~V.} \bibnamefont{Stepantsova}},
  \bibnamefont{and} \bibinfo{author}{\bibfnamefont{V.~M.}
  \bibnamefont{Kaganer}}, \bibinfo{journal}{J. Crystal Growth}
  \textbf{\bibinfo{volume}{118}}, \bibinfo{pages}{163}
  (\bibinfo{year}{1992}{\natexlab{a}}).

\bibitem[{\citenamefont{Charniy
  et~al.}(1992{\natexlab{b}})\citenamefont{Charniy, Morozov, Scherbachov,
  Bublik, and Stepantsova}}]{Charniy1992b}
\bibinfo{author}{\bibfnamefont{L.~A.} \bibnamefont{Charniy}},
  \bibinfo{author}{\bibfnamefont{A.~N.} \bibnamefont{Morozov}},
  \bibinfo{author}{\bibfnamefont{K.~D.} \bibnamefont{Scherbachov}},
  \bibinfo{author}{\bibfnamefont{V.~T.} \bibnamefont{Bublik}},
  \bibnamefont{and} \bibinfo{author}{\bibfnamefont{I.~V.}
  \bibnamefont{Stepantsova}}, \bibinfo{journal}{J. Crystal Growth}
  \textbf{\bibinfo{volume}{116}}, \bibinfo{pages}{369}
  (\bibinfo{year}{1992}{\natexlab{b}}).

\bibitem[{\citenamefont{Fujimoto}(1992)}]{Fujimoto1992}
\bibinfo{author}{\bibfnamefont{I.}~\bibnamefont{Fujimoto}},
  \bibinfo{journal}{Materials Science and Engineering}
  \textbf{\bibinfo{volume}{B14}}, \bibinfo{pages}{426} (\bibinfo{year}{1992}).

\bibitem[{\citenamefont{Nolte}(1999)}]{Nolte1999}
\bibinfo{author}{\bibfnamefont{D.~D.} \bibnamefont{Nolte}},
  \bibinfo{journal}{J. Appl.\ Phys.} \textbf{\bibinfo{volume}{85}},
  \bibinfo{pages}{6259} (\bibinfo{year}{1999}).

\bibitem[{\citenamefont{Yu et~al.}(1992)\citenamefont{Yu, Kaminska, and
  Liliental-Weber}}]{Yu1992}
\bibinfo{author}{\bibfnamefont{K.~M.} \bibnamefont{Yu}},
  \bibinfo{author}{\bibfnamefont{M.}~\bibnamefont{Kaminska}}, \bibnamefont{and}
  \bibinfo{author}{\bibfnamefont{Z.}~\bibnamefont{Liliental-Weber}},
  \bibinfo{journal}{J. Appl.\ Phys.} \textbf{\bibinfo{volume}{72}},
  \bibinfo{pages}{2850} (\bibinfo{year}{1992}).

\bibitem[{\citenamefont{Kaminska et~al.}(1989)\citenamefont{Kaminska,
  Liliental-Weber, Weber, George, Kortright, Smith, Tsaur, and
  Calawa}}]{Kaminska1989}
\bibinfo{author}{\bibfnamefont{M.}~\bibnamefont{Kaminska}},
  \bibinfo{author}{\bibfnamefont{Z.}~\bibnamefont{Liliental-Weber}},
  \bibinfo{author}{\bibfnamefont{E.~R.} \bibnamefont{Weber}},
  \bibinfo{author}{\bibfnamefont{T.}~\bibnamefont{George}},
  \bibinfo{author}{\bibfnamefont{J.~B.} \bibnamefont{Kortright}},
  \bibinfo{author}{\bibfnamefont{F.~W.} \bibnamefont{Smith}},
  \bibinfo{author}{\bibfnamefont{B.-Y.} \bibnamefont{Tsaur}}, \bibnamefont{and}
  \bibinfo{author}{\bibfnamefont{A.~R.} \bibnamefont{Calawa}},
  \bibinfo{journal}{Appl.\ Phys.\ Lett.} \textbf{\bibinfo{volume}{54}},
  \bibinfo{pages}{1881} (\bibinfo{year}{1989}).

\bibitem[{\citenamefont{Liu et~al.}(1995)\citenamefont{Liu, Nishio, Weber,
  Liliental-Weber, and Walukiewicz}}]{Liu1995}
\bibinfo{author}{\bibfnamefont{X.}~\bibnamefont{Liu}},
  \bibinfo{author}{\bibfnamefont{J.}~\bibnamefont{Nishio}},
  \bibinfo{author}{\bibfnamefont{E.~R.} \bibnamefont{Weber}},
  \bibinfo{author}{\bibfnamefont{Z.}~\bibnamefont{Liliental-Weber}},
  \bibnamefont{and}
  \bibinfo{author}{\bibfnamefont{W.}~\bibnamefont{Walukiewicz}},
  \bibinfo{journal}{Appl.\ Phys.\ Lett.} \textbf{\bibinfo{volume}{67}},
  \bibinfo{pages}{279} (\bibinfo{year}{1995}).

\bibitem[{\citenamefont{Feenstra et~al.}(1993)\citenamefont{Feenstra, Woodall,
  and Pettit}}]{Feenstra1993}
\bibinfo{author}{\bibfnamefont{R.~M.} \bibnamefont{Feenstra}},
  \bibinfo{author}{\bibfnamefont{J.~M.} \bibnamefont{Woodall}},
  \bibnamefont{and} \bibinfo{author}{\bibfnamefont{G.~D.}
  \bibnamefont{Pettit}}, \bibinfo{journal}{Phys.\ Rev.\ Lett.}
  \textbf{\bibinfo{volume}{71}}, \bibinfo{pages}{1176} (\bibinfo{year}{1993}).

\bibitem[{\citenamefont{Luysberg et~al.}(1998)\citenamefont{Luysberg, Sohn,
  Prasad, Specht, Liliental-Weber, Weber, Gebauer, and
  Krause-Rehberg}}]{Luysberg1998}
\bibinfo{author}{\bibfnamefont{M.}~\bibnamefont{Luysberg}},
  \bibinfo{author}{\bibfnamefont{H.}~\bibnamefont{Sohn}},
  \bibinfo{author}{\bibfnamefont{A.}~\bibnamefont{Prasad}},
  \bibinfo{author}{\bibfnamefont{P.}~\bibnamefont{Specht}},
  \bibinfo{author}{\bibfnamefont{Z.}~\bibnamefont{Liliental-Weber}},
  \bibinfo{author}{\bibfnamefont{E.~R.} \bibnamefont{Weber}},
  \bibinfo{author}{\bibfnamefont{J.}~\bibnamefont{Gebauer}}, \bibnamefont{and}
  \bibinfo{author}{\bibfnamefont{R.}~\bibnamefont{Krause-Rehberg}},
  \bibinfo{journal}{J. Appl.\ Phys.} \textbf{\bibinfo{volume}{83}},
  \bibinfo{pages}{561} (\bibinfo{year}{1998}).

\bibitem[{\citenamefont{Staab et~al.}(2001)\citenamefont{Staab, Nieminen,
  Gebauer, {Krause-Rehberg}, Luysberg, Haugk, and Frauenheim}}]{Staab2001}
\bibinfo{author}{\bibfnamefont{T.~E.~M.} \bibnamefont{Staab}},
  \bibinfo{author}{\bibfnamefont{R.~M.} \bibnamefont{Nieminen}},
  \bibinfo{author}{\bibfnamefont{J.}~\bibnamefont{Gebauer}},
  \bibinfo{author}{\bibfnamefont{R.}~\bibnamefont{{Krause-Rehberg}}},
  \bibinfo{author}{\bibfnamefont{M.}~\bibnamefont{Luysberg}},
  \bibinfo{author}{\bibfnamefont{M.}~\bibnamefont{Haugk}}, \bibnamefont{and}
  \bibinfo{author}{\bibfnamefont{T.}~\bibnamefont{Frauenheim}},
  \bibinfo{journal}{Phys.\ Rev.\ Lett.} \textbf{\bibinfo{volume}{87}},
  \bibinfo{pages}{045504} (\bibinfo{year}{2001}).

\bibitem[{\citenamefont{Schultz et~al.}(1998)\citenamefont{Schultz, Egger,
  Scholz, Breitenstein, {Go\"sele}, and Tan}}]{Schultz1998}
\bibinfo{author}{\bibfnamefont{M.}~\bibnamefont{Schultz}},
  \bibinfo{author}{\bibfnamefont{U.}~\bibnamefont{Egger}},
  \bibinfo{author}{\bibfnamefont{R.}~\bibnamefont{Scholz}},
  \bibinfo{author}{\bibfnamefont{O.}~\bibnamefont{Breitenstein}},
  \bibinfo{author}{\bibfnamefont{U.}~\bibnamefont{{Go\"sele}}},
  \bibnamefont{and} \bibinfo{author}{\bibfnamefont{T.~Y.} \bibnamefont{Tan}},
  \bibinfo{journal}{J. Appl.\ Phys.} \textbf{\bibinfo{volume}{83}},
  \bibinfo{pages}{5295} (\bibinfo{year}{1998}).

\bibitem[{\citenamefont{Chaldyshev et~al.}(2001)\citenamefont{Chaldyshev, Bert,
  Musikhin, Suvorova, Preobrazhenskii, Putyato, Semyagin, Werner, and
  {G\"osele}}}]{Chaldyshev2001a}
\bibinfo{author}{\bibfnamefont{V.~V.} \bibnamefont{Chaldyshev}},
  \bibinfo{author}{\bibfnamefont{N.~A.} \bibnamefont{Bert}},
  \bibinfo{author}{\bibfnamefont{Y.~G.} \bibnamefont{Musikhin}},
  \bibinfo{author}{\bibfnamefont{A.~A.} \bibnamefont{Suvorova}},
  \bibinfo{author}{\bibfnamefont{V.~V.} \bibnamefont{Preobrazhenskii}},
  \bibinfo{author}{\bibfnamefont{M.~A.} \bibnamefont{Putyato}},
  \bibinfo{author}{\bibfnamefont{B.~R.} \bibnamefont{Semyagin}},
  \bibinfo{author}{\bibfnamefont{P.}~\bibnamefont{Werner}}, \bibnamefont{and}
  \bibinfo{author}{\bibfnamefont{U.}~\bibnamefont{{G\"osele}}},
  \bibinfo{journal}{Appl.\ Phys.\ Lett.} \textbf{\bibinfo{volume}{79}},
  \bibinfo{pages}{1294} (\bibinfo{year}{2001}).

\bibitem[{\citenamefont{Baraff and {Schl\"uter}}(1985)}]{Baraff1985a}
\bibinfo{author}{\bibfnamefont{G.~A.} \bibnamefont{Baraff}} \bibnamefont{and}
  \bibinfo{author}{\bibfnamefont{M.}~\bibnamefont{{Schl\"uter}}},
  \bibinfo{journal}{Phys.\ Rev.\ Lett.} \textbf{\bibinfo{volume}{55}},
  \bibinfo{pages}{1327} (\bibinfo{year}{1985}).

\bibitem[{\citenamefont{Jansen and Sankey}(1989)}]{Jansen1989}
\bibinfo{author}{\bibfnamefont{R.~W.} \bibnamefont{Jansen}} \bibnamefont{and}
  \bibinfo{author}{\bibfnamefont{O.~F.} \bibnamefont{Sankey}},
  \bibinfo{journal}{Phys.\ Rev.\ B} \textbf{\bibinfo{volume}{39}},
  \bibinfo{pages}{3192} (\bibinfo{year}{1989}).

\bibitem[{\citenamefont{Zhang and Northrup}(1991)}]{Zhang1991}
\bibinfo{author}{\bibfnamefont{S.~B.} \bibnamefont{Zhang}} \bibnamefont{and}
  \bibinfo{author}{\bibfnamefont{J.~E.} \bibnamefont{Northrup}},
  \bibinfo{journal}{Phys.\ Rev.\ Lett.} \textbf{\bibinfo{volume}{67}},
  \bibinfo{pages}{2339} (\bibinfo{year}{1991}).

\bibitem[{\citenamefont{Chadi}(1992)}]{Chadi1992}
\bibinfo{author}{\bibfnamefont{D.~J.} \bibnamefont{Chadi}},
  \bibinfo{journal}{Phys.\ Rev.\ B} \textbf{\bibinfo{volume}{46}},
  \bibinfo{pages}{9400} (\bibinfo{year}{1992}).

\bibitem[{\citenamefont{Landman et~al.}(1997)\citenamefont{Landman, Morgan,
  Schick, Papoulias, and Kumar}}]{Landman1997}
\bibinfo{author}{\bibfnamefont{J.~I.} \bibnamefont{Landman}},
  \bibinfo{author}{\bibfnamefont{C.~G.} \bibnamefont{Morgan}},
  \bibinfo{author}{\bibfnamefont{J.~T.} \bibnamefont{Schick}},
  \bibinfo{author}{\bibfnamefont{P.}~\bibnamefont{Papoulias}},
  \bibnamefont{and} \bibinfo{author}{\bibfnamefont{A.}~\bibnamefont{Kumar}},
  \bibinfo{journal}{Phys.\ Rev.\ B} \textbf{\bibinfo{volume}{55}},
  \bibinfo{pages}{15581} (\bibinfo{year}{1997}).

\bibitem[{\citenamefont{Sankey and Niklewski}(1989)}]{Sankey1989}
\bibinfo{author}{\bibfnamefont{O.~F.} \bibnamefont{Sankey}} \bibnamefont{and}
  \bibinfo{author}{\bibfnamefont{D.~J.} \bibnamefont{Niklewski}},
  \bibinfo{journal}{Phys.\ Rev.\ B} \textbf{\bibinfo{volume}{40}},
  \bibinfo{pages}{3979} (\bibinfo{year}{1989}).

\bibitem[{\citenamefont{Tsai et~al.}(1992)\citenamefont{Tsai, Sankey, and
  Dow}}]{Tsai1992}
\bibinfo{author}{\bibfnamefont{M.-H.} \bibnamefont{Tsai}},
  \bibinfo{author}{\bibfnamefont{O.~F.} \bibnamefont{Sankey}},
  \bibnamefont{and} \bibinfo{author}{\bibfnamefont{J.~D.} \bibnamefont{Dow}},
  \bibinfo{journal}{Phys.\ Rev.\ B} \textbf{\bibinfo{volume}{46}},
  \bibinfo{pages}{10464} (\bibinfo{year}{1992}).

\bibitem[{\citenamefont{Chadi and Cohen}(1973)}]{Chadi1973}
\bibinfo{author}{\bibfnamefont{D.~J.} \bibnamefont{Chadi}} \bibnamefont{and}
  \bibinfo{author}{\bibfnamefont{M.~L.} \bibnamefont{Cohen}},
  \bibinfo{journal}{Phys.\ Rev.\ B} \textbf{\bibinfo{volume}{8}},
  \bibinfo{pages}{5747} (\bibinfo{year}{1973}).

\bibitem[{\citenamefont{{P\"oykk\"o} et~al.}(1996)\citenamefont{{P\"oykk\"o},
  Puska, Alatalo, and Nieminen}}]{Poykko1996b}
\bibinfo{author}{\bibfnamefont{S.}~\bibnamefont{{P\"oykk\"o}}},
  \bibinfo{author}{\bibfnamefont{M.~J.} \bibnamefont{Puska}},
  \bibinfo{author}{\bibfnamefont{M.}~\bibnamefont{Alatalo}}, \bibnamefont{and}
  \bibinfo{author}{\bibfnamefont{R.~M.} \bibnamefont{Nieminen}},
  \bibinfo{journal}{Phys.\ Rev.\ B} \textbf{\bibinfo{volume}{54}},
  \bibinfo{pages}{7909} (\bibinfo{year}{1996}).

\bibitem[{\citenamefont{Makov et~al.}(1996)\citenamefont{Makov, Shah, and
  Payne}}]{Makov1996}
\bibinfo{author}{\bibfnamefont{G.}~\bibnamefont{Makov}},
  \bibinfo{author}{\bibfnamefont{R.}~\bibnamefont{Shah}}, \bibnamefont{and}
  \bibinfo{author}{\bibfnamefont{M.~C.} \bibnamefont{Payne}},
  \bibinfo{journal}{Phys.\ Rev.\ B} \textbf{\bibinfo{volume}{53}},
  \bibinfo{pages}{15513} (\bibinfo{year}{1996}).

\bibitem[{\citenamefont{Puska et~al.}(1998)\citenamefont{Puska, {P\"oykk\"o},
  Pesola, and Nieminen}}]{Puska1998}
\bibinfo{author}{\bibfnamefont{M.~J.} \bibnamefont{Puska}},
  \bibinfo{author}{\bibfnamefont{S.}~\bibnamefont{{P\"oykk\"o}}},
  \bibinfo{author}{\bibfnamefont{M.}~\bibnamefont{Pesola}}, \bibnamefont{and}
  \bibinfo{author}{\bibfnamefont{R.~M.} \bibnamefont{Nieminen}},
  \bibinfo{journal}{Phys.\ Rev.\ B} \textbf{\bibinfo{volume}{58}},
  \bibinfo{pages}{1318} (\bibinfo{year}{1998}).

\bibitem[{\citenamefont{Bockstedte et~al.}(1997)\citenamefont{Bockstedte, Kley,
  Neugebauer, and Scheffler}}]{Bockstedte1997}
\bibinfo{author}{\bibfnamefont{M.}~\bibnamefont{Bockstedte}},
  \bibinfo{author}{\bibfnamefont{A.}~\bibnamefont{Kley}},
  \bibinfo{author}{\bibfnamefont{J.}~\bibnamefont{Neugebauer}},
  \bibnamefont{and}
  \bibinfo{author}{\bibfnamefont{M.}~\bibnamefont{Scheffler}},
  \bibinfo{journal}{Comput.\ Phys.\ Commun.} \textbf{\bibinfo{volume}{107}},
  \bibinfo{pages}{187} (\bibinfo{year}{1997}).

\bibitem[{\citenamefont{Hohenberg and Kohn}(1964)}]{Hohenberg1964}
\bibinfo{author}{\bibfnamefont{P.}~\bibnamefont{Hohenberg}} \bibnamefont{and}
  \bibinfo{author}{\bibfnamefont{W.}~\bibnamefont{Kohn}},
  \bibinfo{journal}{Phys.\ Rev.} \textbf{\bibinfo{volume}{136}},
  \bibinfo{pages}{B864} (\bibinfo{year}{1964}).

\bibitem[{\citenamefont{Ceperley and Alder}(1980)}]{Ceperley1980}
\bibinfo{author}{\bibfnamefont{D.~M.} \bibnamefont{Ceperley}} \bibnamefont{and}
  \bibinfo{author}{\bibfnamefont{G.~J.} \bibnamefont{Alder}},
  \bibinfo{journal}{Phys.\ Rev.\ Lett.} \textbf{\bibinfo{volume}{45}},
  \bibinfo{pages}{566} (\bibinfo{year}{1980}).

\bibitem[{\citenamefont{Perdew and Zunger}(1981)}]{Perdew1981}
\bibinfo{author}{\bibfnamefont{J.}~\bibnamefont{Perdew}} \bibnamefont{and}
  \bibinfo{author}{\bibfnamefont{A.}~\bibnamefont{Zunger}},
  \bibinfo{journal}{Phys.\ Rev.\ B} \textbf{\bibinfo{volume}{23}},
  \bibinfo{pages}{5048} (\bibinfo{year}{1981}).

\bibitem[{\citenamefont{Kleinman and Bylander}(1982)}]{Kleinman1982}
\bibinfo{author}{\bibfnamefont{L.}~\bibnamefont{Kleinman}} \bibnamefont{and}
  \bibinfo{author}{\bibfnamefont{D.~M.} \bibnamefont{Bylander}},
  \bibinfo{journal}{Phys.\ Rev.\ Lett.} \textbf{\bibinfo{volume}{48}},
  \bibinfo{pages}{1425} (\bibinfo{year}{1982}).

\bibitem[{\citenamefont{Hamann}(1989)}]{Hamann1989}
\bibinfo{author}{\bibfnamefont{D.~R.} \bibnamefont{Hamann}},
  \bibinfo{journal}{Phys.\ Rev.\ B} \textbf{\bibinfo{volume}{40}},
  \bibinfo{pages}{2980} (\bibinfo{year}{1989}).

\bibitem[{\citenamefont{Kleinman}(1981)}]{Kleinman1981}
\bibinfo{author}{\bibfnamefont{L.}~\bibnamefont{Kleinman}},
  \bibinfo{journal}{Phys.\ Rev.\ B} \textbf{\bibinfo{volume}{24}},
  \bibinfo{pages}{7412} (\bibinfo{year}{1981}).

\bibitem[{\citenamefont{Kohan et~al.}(2000)\citenamefont{Kohan, Ceder, Morgan,
  and de~Walle}}]{Kohan2000}
\bibinfo{author}{\bibfnamefont{A.~F.} \bibnamefont{Kohan}},
  \bibinfo{author}{\bibfnamefont{G.}~\bibnamefont{Ceder}},
  \bibinfo{author}{\bibfnamefont{D.}~\bibnamefont{Morgan}}, \bibnamefont{and}
  \bibinfo{author}{\bibfnamefont{C.~G.~V.} \bibnamefont{de~Walle}},
  \bibinfo{journal}{Phys.\ Rev.\ B} \textbf{\bibinfo{volume}{61}},
  \bibinfo{pages}{15019} (\bibinfo{year}{2000}).

\bibitem[{\citenamefont{Baraff and {Schl\"uter}}(1984)}]{Baraff1984}
\bibinfo{author}{\bibfnamefont{G.~A.} \bibnamefont{Baraff}} \bibnamefont{and}
  \bibinfo{author}{\bibfnamefont{M.}~\bibnamefont{{Schl\"uter}}},
  \bibinfo{journal}{Phys.\ Rev.\ B} \textbf{\bibinfo{volume}{30}},
  \bibinfo{pages}{1853} (\bibinfo{year}{1984}).

\bibitem[{\citenamefont{Johnson and Ashcroft}(1998)}]{Johnson1998a}
\bibinfo{author}{\bibfnamefont{K.~A.} \bibnamefont{Johnson}} \bibnamefont{and}
  \bibinfo{author}{\bibfnamefont{N.~W.} \bibnamefont{Ashcroft}},
  \bibinfo{journal}{Phys.\ Rev.\ B} \textbf{\bibinfo{volume}{58}},
  \bibinfo{pages}{15548} (\bibinfo{year}{1998}).

\bibitem[{\citenamefont{Hedin}(1965)}]{Hedin1965a}
\bibinfo{author}{\bibfnamefont{L.}~\bibnamefont{Hedin}},
  \bibinfo{journal}{Phys.\ Rev.} \textbf{\bibinfo{volume}{139}},
  \bibinfo{pages}{A796} (\bibinfo{year}{1965}).

\bibitem[{\citenamefont{Hedin and Lundqvist}(1969)}]{Hedin1969a}
\bibinfo{author}{\bibfnamefont{L.}~\bibnamefont{Hedin}} \bibnamefont{and}
  \bibinfo{author}{\bibfnamefont{S.}~\bibnamefont{Lundqvist}},
  \emph{\bibinfo{title}{Solid State Phys.{\em , v. 23}}}
  (\bibinfo{publisher}{Academic Press}, \bibinfo{address}{New York},
  \bibinfo{year}{1969}), p.~\bibinfo{pages}{1}.

\bibitem[{\citenamefont{Hybertsen and Louie}(1986)}]{Hybertsen1986a}
\bibinfo{author}{\bibfnamefont{M.~S.} \bibnamefont{Hybertsen}}
  \bibnamefont{and} \bibinfo{author}{\bibfnamefont{S.~G.} \bibnamefont{Louie}},
  \bibinfo{journal}{Phys.\ Rev.\ B} \textbf{\bibinfo{volume}{34}},
  \bibinfo{pages}{5390} (\bibinfo{year}{1986}).

\bibitem[{\citenamefont{Godby et~al.}(1988)\citenamefont{Godby, {Schl\"uter},
  and Sham}}]{Godby1988a}
\bibinfo{author}{\bibfnamefont{R.~W.} \bibnamefont{Godby}},
  \bibinfo{author}{\bibfnamefont{M.}~\bibnamefont{{Schl\"uter}}},
  \bibnamefont{and} \bibinfo{author}{\bibfnamefont{L.~J.} \bibnamefont{Sham}},
  \bibinfo{journal}{Phys.\ Rev.\ B} \textbf{\bibinfo{volume}{37}},
  \bibinfo{pages}{10159} (\bibinfo{year}{1988}).

\bibitem[{\citenamefont{Rohlfing et~al.}(1993)\citenamefont{Rohlfing,
  {Kr\"uger}, and Pollman}}]{Rohlfing1993a}
\bibinfo{author}{\bibfnamefont{M.}~\bibnamefont{Rohlfing}},
  \bibinfo{author}{\bibfnamefont{P.}~\bibnamefont{{Kr\"uger}}},
  \bibnamefont{and} \bibinfo{author}{\bibfnamefont{J.}~\bibnamefont{Pollman}},
  \bibinfo{journal}{Phys.\ Rev.\ B} \textbf{\bibinfo{volume}{48}},
  \bibinfo{pages}{17791} (\bibinfo{year}{1993}).

\bibitem[{\citenamefont{Monkhorst and Pack}(1976)}]{Monkhorst1976}
\bibinfo{author}{\bibfnamefont{H.~J.} \bibnamefont{Monkhorst}}
  \bibnamefont{and} \bibinfo{author}{\bibfnamefont{J.~D.} \bibnamefont{Pack}},
  \bibinfo{journal}{Phys.\ Rev.\ B} \textbf{\bibinfo{volume}{13}},
  \bibinfo{pages}{5188} (\bibinfo{year}{1976}).

\bibitem[{\citenamefont{Chadi et~al.}(1997)\citenamefont{Chadi, Citrin, Park,
  Adler, Marcus, and Gossmann}}]{Chadi1997}
\bibinfo{author}{\bibfnamefont{D.~J.} \bibnamefont{Chadi}},
  \bibinfo{author}{\bibfnamefont{P.~H.} \bibnamefont{Citrin}},
  \bibinfo{author}{\bibfnamefont{C.~H.} \bibnamefont{Park}},
  \bibinfo{author}{\bibfnamefont{D.~L.} \bibnamefont{Adler}},
  \bibinfo{author}{\bibfnamefont{M.~A.} \bibnamefont{Marcus}},
  \bibnamefont{and} \bibinfo{author}{\bibfnamefont{H.-J.}
  \bibnamefont{Gossmann}}, \bibinfo{journal}{Phys.\ Rev.\ Lett.}
  \textbf{\bibinfo{volume}{79}}, \bibinfo{pages}{4834} (\bibinfo{year}{1997}).

\bibitem[{\citenamefont{{Bl\"ochl} et~al.}(1993)\citenamefont{{Bl\"ochl},
  Smargiassi, Car, Laks, Andreoni, and Pantelides}}]{Blochl1993}
\bibinfo{author}{\bibfnamefont{P.~E.} \bibnamefont{{Bl\"ochl}}},
  \bibinfo{author}{\bibfnamefont{E.}~\bibnamefont{Smargiassi}},
  \bibinfo{author}{\bibfnamefont{R.}~\bibnamefont{Car}},
  \bibinfo{author}{\bibfnamefont{D.~B.} \bibnamefont{Laks}},
  \bibinfo{author}{\bibfnamefont{W.}~\bibnamefont{Andreoni}}, \bibnamefont{and}
  \bibinfo{author}{\bibfnamefont{S.~T.} \bibnamefont{Pantelides}},
  \bibinfo{journal}{Phys.\ Rev.\ Lett.} \textbf{\bibinfo{volume}{70}},
  \bibinfo{pages}{2435} (\bibinfo{year}{1993}).

\bibitem[{\citenamefont{Sun et~al.}(1992)\citenamefont{Sun, Watkins, Rong,
  Fotiadis, and Poindexter}}]{Sun1992}
\bibinfo{author}{\bibfnamefont{H.-J.} \bibnamefont{Sun}},
  \bibinfo{author}{\bibfnamefont{G.~D.} \bibnamefont{Watkins}},
  \bibinfo{author}{\bibfnamefont{F.~C.} \bibnamefont{Rong}},
  \bibinfo{author}{\bibfnamefont{L.}~\bibnamefont{Fotiadis}}, \bibnamefont{and}
  \bibinfo{author}{\bibfnamefont{E.~H.} \bibnamefont{Poindexter}},
  \bibinfo{journal}{Appl.\ Phys.\ Lett.} \textbf{\bibinfo{volume}{60}},
  \bibinfo{pages}{718} (\bibinfo{year}{1992}).

\bibitem[{\citenamefont{Weber et~al.}(1982)\citenamefont{Weber, Ennen,
  Kaufmann, Windscheif, Schneider, and Wosinski}}]{Weber1982a}
\bibinfo{author}{\bibfnamefont{E.~R.} \bibnamefont{Weber}},
  \bibinfo{author}{\bibfnamefont{H.}~\bibnamefont{Ennen}},
  \bibinfo{author}{\bibfnamefont{U.}~\bibnamefont{Kaufmann}},
  \bibinfo{author}{\bibfnamefont{J.}~\bibnamefont{Windscheif}},
  \bibinfo{author}{\bibfnamefont{J.}~\bibnamefont{Schneider}},
  \bibnamefont{and} \bibinfo{author}{\bibfnamefont{T.}~\bibnamefont{Wosinski}},
  \bibinfo{journal}{J. Appl.\ Phys.} \textbf{\bibinfo{volume}{53}},
  \bibinfo{pages}{6140} (\bibinfo{year}{1982}).

\end{thebibliography}
\end{document}